\documentclass{jfm_arxiv}

\usepackage{graphicx}
\usepackage{newtxtext}
\usepackage{newtxmath}
\usepackage{natbib}
\usepackage{siunitx}
\usepackage{physics}
\AtBeginDocument{\RenewCommandCopy\qty\SI}
\usepackage{hyperref}
\usepackage{cleveref}
\hypersetup{
colorlinks = true,
urlcolor   = blue,
citecolor  = black,
}
\newcommand{\RomanNumeralCaps}[1]
\linenumbers

\DeclareSIUnit{\pixel}{px}
\newcommand{\kindex}[2]{\ensuremath{{#1}_{\scalebox{0.65}{#2}}}}
\newcommand{\U}{\textrm{U}}
\newcommand{\Uinf}{\kindex{\U}{$\infty$}}

\newcommand{\adottext}{\kindex{\dot{\alpha}}{ss}c/\Uinf}

\newcommand{\Clqs}{\kindex{C}{l,qs}}

\graphicspath{{../figurematter/latex/figs/}}
\usepackage{marvosym,graphicx,setspace,eurosym,xcolor}

\title{Dynamic stall reattachment revisited}

\author{Sahar Rezapour \and Karen Mulleners \corresp{\email{karen.mulleners@epfl.ch}}}

\affiliation{Institute of Mechanical Engineering,
École Polytechnique Fédérale de Lausanne (EPFL),
Lausanne,
Switzerland
}

\begin{document}
\maketitle
	
\begin{abstract}
Dynamic stall on airfoils is an undesirable and potentially dangerous phenomenon. The motto for aerodynamic systems with unsteadily moving wings, such as helicopters or wind turbines, is that prevention beats recovery. In case prevention fails or is not feasible, we need to know when recovery starts, how long it takes, and how we can improve it. This study revisits dynamic stall reattachment to identify the sequence of events during flow and load recovery and to characterise key observable features in the pressure, force, and flow field. Our analysis is based on time-resolved velocity field and surface pressure data obtained experimentally for a two-dimensional, sinusoidally pitching thin airfoil. Stall recovery is a transient process that does not start immediately when the angle of attack falls below the critical stall angle. The onset of recovery is delayed to angles below the critical stall angle and the duration of the reattachment delay decreases with increasing unsteadiness of the pitching motion. An angle of attack below the critical angle is a necessary, but not sufficient condition to initiate the stall recovery process. We identified a critical value of the leading-edge suction parameter, independent of the pitch rate, that is a threshold beyond which reattachment consistently initiates. Based on prominent changes in the evolution of the shear layer, the leading-edge suction, and the lift deficit due to stall, we divided the reattachment process into three stages: the reaction delay, wave propagation, and the relaxation stage, and extracted the characteristic features and time-scales for each stage. 
\end{abstract}



\section{Introduction}
Dynamic stall happens when the angle of attack of an airfoil exceeds its critical stall angle, either due to unsteady airfoil motions or variations in flow conditions, such as gust encounters \citep{McCroskey1981,Jones2021}.
The dynamic stall phenomenon is commonly observed on the retreating blades of helicopter rotors in forward flight, horizontal- and vertical-axis wind turbines, cross-flow hydro-turbines, and micro-aerial vehicles \citep{Buchner2018, dave.2023, Santos2022, LeFouest2022a}. 
A typical dynamic stall cycle includes attached flow, the emergence and spreading of flow reversal on the airfoil suction side, the formation of a large-scale dynamic stall vortex, the separation of the first dynamic stall vortex initiating full stall, massive flow separation, and eventually flow reattachment \citep{Carr1977,Shih1992,Mulleners2013}.  
Dynamic stall is affected by the airfoil geometry, the wing kinematics, and flow conditions \citep{Visbal2018, Corke2015, Choudhry2014a}. 
The airfoil geometry mainly affects the stall onset and hysteresis loops.
Thicker airfoils generally stall at higher angles of attack and have larger hysteresis loops than thinner airfoils. 

The stall onset and flow recovery are delayed in dynamic stall compared to classic static airfoil separation and reattachment \citep{Carr1977,Shih1992,LeFouest2021}. 
The separation delay is well-characterised using the non-dimensional measure of the kinematic unsteadiness defined as the non-dimensional pitch rate \citep{Mulleners2012,Ayancik2022,Kiefer2022}. 
The delay in stall onset increases the maximum attainable lift, as the lift continues to grow with increasing angle of attack during stall development.
Although the initial stall delay and lift overshoot may seem advantageous, they lead to large, unsteady aerodynamic loads that reduce efficiency, introduce strong vibrations, and increase structural stress.

Due to the adverse effects of dynamic stall onset, this phase has historically received more attention than stall recovery \citep{Sheng:2008fz, Morris:2013cl, Mulleners2012} and the developed dynamic stall models typically predict separation onset more accurately than the recovery onset \citep{Sheng:2008fz, Ayancik2022, Damiola2024}. 
A more profound understanding of the reattachment process is essential to improve the modelling of the overall aerodynamic performance, and to develop strategies for gust exits and to alleviate dynamic hysteresis. 
The hysteresis arises directly from the stall onset and recovery delays, and the amount of hysteresis is influenced by factors such as the degree of flow separation, the reduced frequency of the pitching motion, and the presence of a strong dynamic stall vortex~\citep{Ekaterinaris:1998gx, Williams2015}.
In the context of gust-induced flow separation, considerable research has focused on exploring the aerodynamic response of wings to gusts, developing mitigation strategies, and modelling the aerodynamic performance of wings \citep{Perrotta2017, Sedky2020, Jones2021}. 
Resulting gust-response models and control strategies yield better results during gust entry than during the gust exit \citep{andreu-angulo.2023, gementzopoulos.2024}. 

Among the first researchers to systematically study the unsteady flow reattachment process on airfoils were \citet{Niven:1989uq}.
They considered experimental data from various airfoils undergoing ramp-down motions conducted in the facilities of Glasgow University.
The results of these experimental campaigns were collected into the University of Glasgow's aerofoil database \citep{Galbraith:1992ws}.
These early studies found that flow reattachment is always initiated at an angle of attack close to the static stall and progresses from the leading toward the trailing-edge \citep{Galbraith:1992ws, Niven:1989uq}.
\citet{Green:1995bq} were the first to identify a ramp-down wave travelling from the leading to the trailing-edge as an essential part of the recovery process. 
Through smoke flow visualisation, the ramp-down wave was observed to convect excess wake fluid past the airfoil, which is an essential precursor for actual boundary layer reattachment. 
The footprint of the ramp-down wave in the surface pressure data is a local ridge indicating a negative rate of change in suction that travels from the leading to the trailing edge.
The travelling speed of the ramp-down wave is independent of the reduced pitch rate and airfoil profile and is significantly faster than the actual reattachment.
The actual reattachment is indicated by the local recovery of suction, which moves with a rate that strongly depends on the pitch rate.

The reattachment process on periodically pitching airfoils is more complicated than in simple ramp-down motions due to the influence of the strong transient vortex shedding in the near wake.
The issue was tackled by \citet{Ahmed1994} for a large amplitude sinusoidal pitching motion based on flow field visualisation.
Quantitative velocity and density field information was obtained using laser Doppler velocimetry (LDV) and point diffraction interferometry.
Reattachment was again found to begin near the static stall angle of attack, but the point-wise character of the laser Doppler velocimetry technique did not allow for a deeper insight into the process itself. 
In this paper, we revisit the reattachment of dynamically stalled flows on oscillating airfoils. We combine flow field measurements from time-resolved particle image velocimetry with surface pressure measurements from a thin airfoil in deep stall conditions. 

From the surface pressure data, we can extract the leading-edge suction parameter, which quantifies the suction force at an airfoil leading-edge.
The leading-edge suction parameter is a reliable indicator of the onset and evolution of dynamic stall \citep{Ramesh2013, Ramesh2014, Deparday2019, Miotto2022}. 
For a given airfoil shape and Reynolds number, there is a critical threshold beyond which leading edge vortex formation initiates \citep{ramesh.2018}. 
The critical leading-edge suction parameter value remains largely independent of the motion kinematics, except when high degrees of trailing-edge flow separation are present. 
Between the start of the formation and the shedding of the leading edge vortex, the leading-edge suction parameter increases beyond its critical value. 
The maximum leading-edge suction parameter increases with increasing pitch rate of the motion and sharply drops to near-zero values after leading-edge flow separation occurs \citep{Deparday2019, Narsipur2020, Sudharsan2023}.
The leading-edge suction parameter effectively captures the events at the very leading edge, and it is less sensitive to post-stall vortex shedding events occurring away from the leading edge \citep{Sudharsan2024}.
As dynamic stall reattachment starts from the leading edge, the leading-edge suction parameter has the potential to be an insightful quantity that we analyse in this paper as part of our revisit of dynamic stall reattachment. 

\section{Experimental material and methods}
Wind tunnel experiments were conducted to investigate the dynamic stall life cycle on a constantly pitching airfoil in a uniform flow at a free-stream Reynolds number $Re=\num{9.2e5}$ based on the chord length $c$ (Mach number $Ma=\num{0.14}$).
The experiments were conducted in the closed-circuit, low-speed wind tunnel at the German Aerospace Centre (DLR) in G\"{o}ttingen. 
The wind tunnel had an open test section of \qty{1.3}{\meter} length and a rectangular nozzle of \qtyproduct{0.75x1.05}{\meter}.

A two-dimensional airfoil model with an OA209 profile was used for the experiments. 
The airfoil had a maximum thickness-to-chord ratio of $\qty{9}{\percent}$, a chord length of \qty{0.3}{\meter}, and an aspect ratio of \num{5}. 
The static stall angle of this airfoil under the given experimental conditions was $\kindex{\alpha}{ss}=\ang{21.4}$, and was determined from a static polar \citep{Mulleners2012}.
The airfoil was subjected to a sinusoidally oscillating motion about its quarter chord axis with a mean incidence $\kindex{\alpha}{0}$, an amplitude $\kindex{\alpha}{1}$, and an oscillation frequency $\kindex{f}{osc}$.
The latter is preferably written in dimensionless form as the reduced frequency $k=\pi\,\kindex{f}{osc}\,c/\Uinf$, where $\Uinf$ is the free-stream velocity.
The mean incidence, amplitude and reduced frequency were varied such that
$\kindex{\alpha}{0} \in \{\ang{18}, \ang{20}, \ang{22}\}$,
$\kindex{\alpha}{1} \in  \{\ang{6}, \ang{8}\}$, and
$k \in \{\num{0.050}, \num{0.075}, \num{0.10}\}$.
In this study, we focus only on cases that resulted in deep dynamics stall, which according to \cite{Mulleners2012} are cases where dynamic stall occurs before the airfoil reaches the maximum angle of attack. 
The data has been used in previous studies \citep{Mulleners2012,Mulleners2013,Ansell:2019bw,Deparday2019,Ayancik2022}.

The surface pressure distribution was recorded using \num{41} differential pressure transducers (type Kulite XCQ-093) mounted along the central cross-sectional plane of the airfoil (see \cite{Mulleners2013} or \cref{fig:LEarea} for the location of the pressure sensors).
The pressure data were sampled at a rate of \qty{6}{\kilo\hertz} for a duration of \qty{15}{s}, which corresponds to approximately \num{80} cycles for the highest pitching frequency and \num{40} for the lowest frequency. 
The airfoil surface pressure distributions were integrated to obtain the lift and moment coefficients.
The leading-edge suction parameter is evaluated following the procedure explained in \citet{Deparday2019}.
The procedure involves determining the leading-edge suction vector by integrating pressure data from \num{13} unsteady pressure sensors positioned within the front \qty{10}{\percent} of the airfoil.
The experimental leading-edge suction parameter, \kindex{A}{0}, is then obtained by projecting the leading-edge suction vector along the chord-wise direction, following the approach of \citet{Katz_Plotkin_2001}.

Stereoscopic time-resolved particle image velocimetry (TR-PIV) was conducted in the cross-sectional plane at the model mid-span.
The TR-PIV system consisted of a diode-pumped Nd:YAG laser (Lee Laser, LDP-$200$MQG Dual) that emitted laser pulses with an energy of approximately \qty{10}{\milli\joule} per pulse @ \qty{3}{\kilo\hertz} and two CMOS cameras (Photron Ultima APX-RS). 
The vertical plane at model mid-span was illuminated by the laser from above and the cameras were mounted in a stereoscopic set-up alongside the wind tunnel diffuser. 
The width of the field of view covered the entire chord for the relevant angle of attack range.
Time series of \num{3072} images pairs at full camera resolution (\qtyproduct{1024x1024}{\pixel}) were recorded at \qty{1500}{Hz}.
The time delay between the laser pulses in the image pairs was \qty{30}{\micro\second}.
The camera buffers allowed us to record images for \qty{2}{\second}, covering \num{5} full oscillation cycles for the lowest pitching frequency and up to \num{10} cycles for the highest frequency.  
After mapping the views of both cameras, the dimensions of the PIV measurement window were \qtyproduct{335x165}{\milli\meter} with a spatial resolution of \qty{5.0}{\pixel\per\milli\meter}.
The PIV images were processed using an interrogation window size of \qtyproduct{32x32}{\pixel} and an overlap of approximately \qty{80}{\percent} yielding a grid spacing of \qty{6}{\pixel} or \qty{1.2}{\milli\meter} which is less than \num{0.005}\,$c$.
The interrogation window size was minimised, ensuring an acceptable signal-to-noise ratio.
The window overlap on the other hand was maximised to avoid artificial smoothing of velocity gradients \citep{Richard2006}.
The velocity fields were rotated into the airfoil reference system with the $x$-axis along the chord, the $y$-axis along the span, and the $z$-axis upward perpendicular to the chord.
The origin is located at the rotation axis at the airfoil quarter chord axis.
Simultaneously to the TR-PIV, the surface pressure distribution at the model mid-span was scanned at approximately \qty{6}{\kilo\hertz} for about \qty{15}{\second}.
The data acquisition was synchronised with the recording of the PIV images, allowing for straightforward assignment of the instantaneous pressure distributions to each of the acquired velocity fields.
Additional details about the experimental set-up and measurements can be found in \citet{Mulleners2010, Mulleners2012,Mulleners2013}. 

The finite-time Lyapunov exponent (FTLE) is calculated using the time-resolved flow field data to extract and analyse separation and reattachment lines and to identify Lagrangian coherent flow structures \citep{Green2011}. 
The velocity field is artificially seeded and integrated backward in time to obtain negative FTLE ridges (nFTLE), and forward in time to obtain the positive FTLE ridges (pFTLE).
The backward integration is similar to using smoke visualisation and shows where the particles come from.
The nFTLE ridges indicate regions where nearby flow particles experience the highest attraction, such as near separation lines, and the pFTLE ridges indicate regions where nearby flow particles are repelled \citep{Haller2002,Shadden2005}.
The intersection of nFTLE and pFTLE ridges indicates the location of a saddle point.
Monitoring the emergence and trajectories of saddle points provides insights into the timing and location of vortex formation, and can help in understanding flow separation and attachment \citep{Mulleners2012, rockwood.2018, Kissing2020}.

\section{Results}
Here, we revisit the reattachment process of a thin airfoil undergoing deep dynamic stall based on time-resolved pressure and velocity field measurements. 
First, we revise the overall characteristics of dynamic stall by the example of a selected representative pitching cycle.
Then, we focus on the flow development during reattachment, identify successive reattachment stages, and analyse the corresponding characteristic surface footprints.
Finally, we will establish a critical condition for the onset of stall recovery, and quantify how the time scales associated with the different reattachment stages evolve as a function of the effective unsteadiness of the pitching motion.
A representative cycle is used for the detailed discussion, and the analysis is then extended to multiple cycles of various pitching kinematics that all lead to deep stall.  

\subsection{Footprints of dynamic stall}

\begin{figure}
\includegraphics{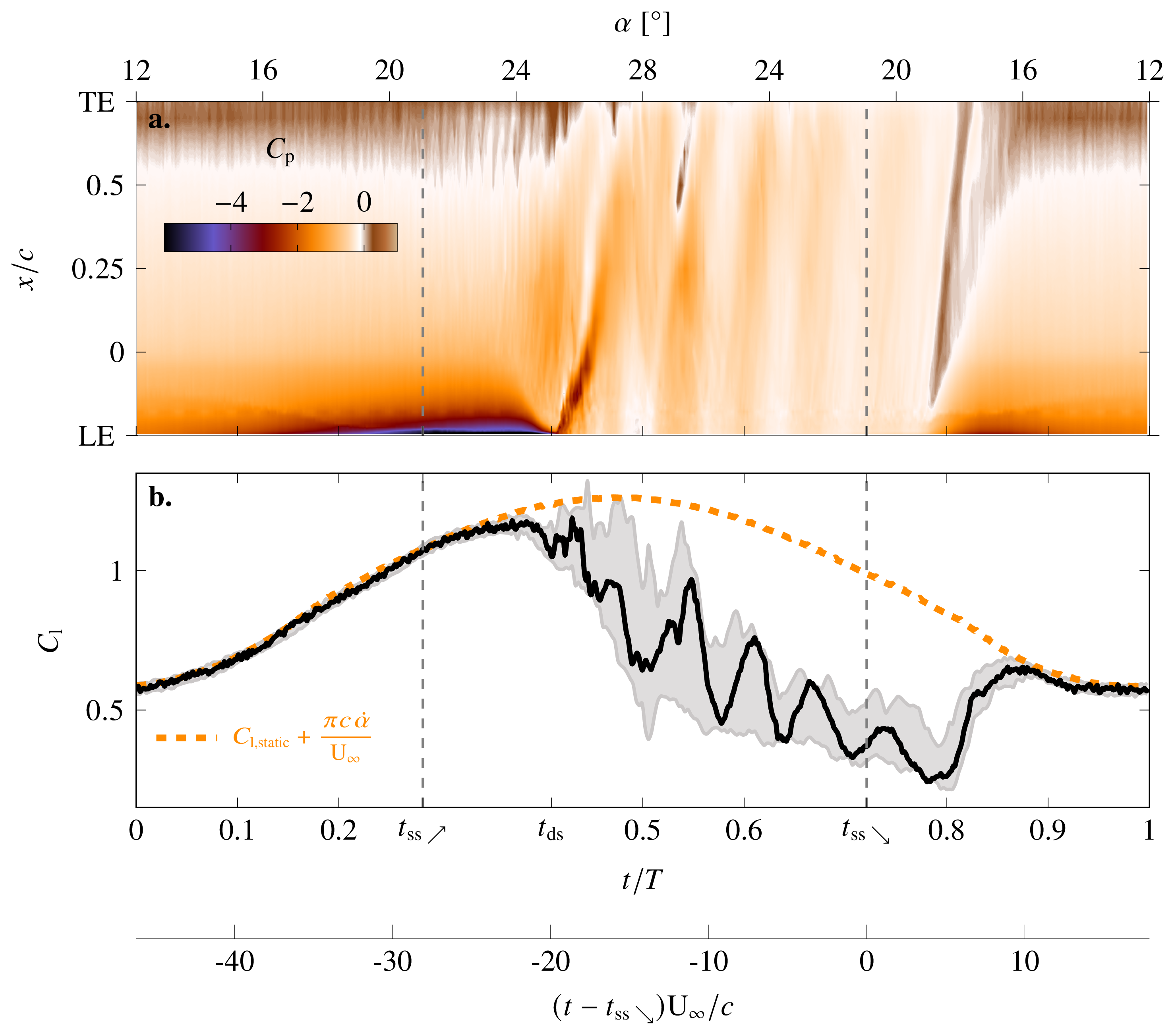}
\caption{
Spatio-temporal evolution of the suction-side surface pressure coefficient (a) and temporal evolution of the lift coefficient (b) for a selected pitching cycle indicated by the solid black line ($\kindex{\alpha}{0}=\ang{20}$, $\kindex{\alpha}{1}=\ang{8}$, $k=\num{0.05}$, $\kindex{\dot{\alpha}}{ss}=\num{0.0135}$).
The shaded gray bands in the lift evolution represent the area between the minimum and maximum envelopes obtained from \num{39} recorded cycles.
The thick dashed line shows the quasi-static evolution of the lift coefficient $\kindex{C}{l,qs}$.
Vertical dashed lines indicate the moment the static stall angle is exceeded during pitch-up ($\kindex{t}{ss $\nearrow$}$) and the moment the angle of attack falls below the static stall angle during pitch-down ($\kindex{t}{ss $\searrow$}$). 
The extra axis on top indicated the angle of attack variation for the cycle.
The extra axis below indicated the non-dimenional time variation shifted based on the instant the geometric angle of attack falls below the critical static stall angle during the pitch-down motion ($\kindex{t}{ss $\searrow$}$).}
\label{fig:forcepanel}
\end{figure}

The typical surface pressure and force response of a single representative cycle of a continuously sinusoidally pitching airfoil undergoing deep dynamic stall is presented in \cref{fig:forcepanel}. 
The data were obtained by oscillating the airfoil around a mean angle of $\kindex{\alpha}{0}=\ang{20}$, with an amplitude $\kindex{\alpha}{1}=\ang{8}$, and a reduced frequency of $k=\num{0.05}$.
We take the start of the cycle at the moment when the airfoil angle of attack is lowest and the pitch-up motion begins.
Additional cycles of the lift response are presented in \cref{fig:more_cycles} in \cref{appendix:morecycles}. 

At the beginning of the cycle, the flow is attached and the lift coefficient increases with the angle of attack, which is primarily due to an increase in the leading edge suction (\cref{fig:forcepanel}a). 
The measured lift coefficient follows the quasi-static lift response prediction for an airfoil pitching around the quarter-chord axis:
\begin{align}
\Clqs(t) &= \kindex{C}{l,static}(\alpha(t)) + 2\pi\frac{\dot{\alpha}(t)c}{2\Uinf} \label{eqn:qslift}\\ 
& = \kindex{C}{l,static}(\alpha(t)) + 2\pi \Delta\alpha(t) \quad, \label{eqn:qsliftda}
\end{align}
with $\kindex{C}{l,static}(\alpha(t))$ the linear extrapolation of the static lift response for attached flow and $\dot{\alpha}(t)$ the instantaneous pitch rate (\cref{fig:forcepanel}b).
The contribution of the unsteady pitching motion to the lift can also be expressed as the result of a variation in the effective angle of attack \kindex{\alpha}{eff}, by $\Delta\alpha=\kindex{\alpha}{eff}-\alpha$ (\cref{appendix:alphaeff}).
When the airfoil is pitching up, the effective angle of attack is increased by $\dot{\alpha}(t)c/(2\Uinf)$, and we expect an increase in the lift coefficient compared to the static lift response if the flow is attached. 
When the airfoil is pitching down, the effective angle of attack is decreased by $\dot{\alpha}(t)c/(2\Uinf)$. 
As the variations $\Delta\alpha$ will be typically less than \ang{1.6} for the kinematics considered here (\cref{appendix:alphaeff}), we use the geometric angle of attack as our reference.

For the example pitching motion presented in \cref{fig:forcepanel}, the static stall angle of attack is reached at $t/T=0.28$, but the leading edge suction and the lift coefficient continue to increase until $t/T=0.40$, when dynamic stall occurs.
The onset of dynamic stall is defined as the detachment of the primary stall vortex, which coincides with a decrease in lift and the breakdown of the leading edge suction. 
The detachment of the stall vortex is marked by the emergence of a saddle point near the leading edge \citep{Mulleners2012}. 
The onset of dynamic stall is indicated by the axis tick label $\kindex{t}{ds}$ in \cref{fig:forcepanel} and occurs after the static stall angle is exceeded, but before the maximum angle of attack is reached at $t/T=0.5$, classifying this case as a typical deep stall case.
The delayed onset of stall and the associated lift overshoot are key features of dynamic stall.
The onset and development of dynamic stall during the pitch-up half of the cycle has been discussed in detail in previous work based on the same data set used in this paper \citep{Mulleners2012,Mulleners2013,Ansell:2019bw,Deparday2019}.

\begin{figure}
\includegraphics[scale=1]{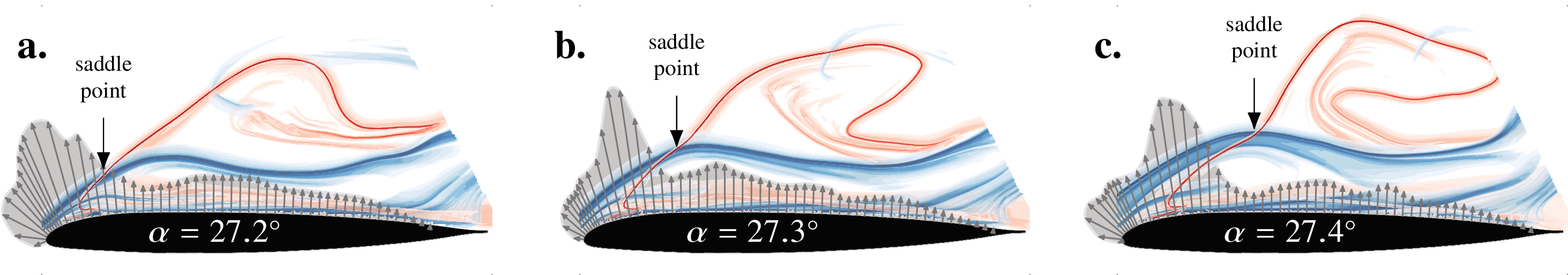}
\caption{Combined visualisation of the instantaneous chord wise surface pressure distribution on the suction side and the nFTLE and pFTLE ridges for three selected time instants immediately following dynamic stall onset for the sinusoidal pitching motion presented in \cref{fig:forcepanel} ($\alpha=\ang{27.2}$ (a), $\alpha=\ang{27.3}$ (b), $\alpha=\ang{27.4}$ (c)). 
The pressure distribution is visualised by arrows normal to the surface, where the length of arrow indicates the magnitude of the pressure coefficient.
Only negative pressure coefficients are displayed. 
The intersection of the nFTLE (red) and pFTLE (blue) ridges indicates the location of a saddle point.}
\label{fig:saddles}
\end{figure}

From here on, we focus on the post-stall behaviour and the stall reattachment process.
A distinct post-stall feature in the space-time representation of the surface pressure distribution is the footprint of the primary dynamic stall vortex in the form of a local minimum pressure trace (\cref{fig:forcepanel}a).
This low-pressure trace originates at the leading edge after dynamic stall onset and moves towards the trailing edge between $t/T=\numrange{0.40}{0.50}$.
Interestingly, the chord-wise location of the local pressure minimum does not correspond to the core of the dynamic stall vortex, but to the upstream saddle point that marks the separation of the stall vortex from the feeding shear layer (\cref{fig:saddles}).
Similar behaviour was observed by \citet{rockwood.2018} for the flow around a circular cylinder.
The location of the saddle point was determined as the intersection of the negative and positive finite time Lyapunov exponent (nFTLE and pFTLE) ridges. 

Dynamic stall onset marks the start of the fully stalled stage of the dynamic stall cycle, which is characterised by the repeated formation and shedding of large scale coherent stall vortices.
This post-stall vortex shedding creates additional low-pressure traces in the space-time representation of the surface pressure distribution and oscillations in the evolution of the instantaneous lift coefficient. 
The surface pressure traces and lift oscillations emerge approximately every \numrange{4}{5} convective times, where the convective time is defined as $\kindex{t}{c}=\Uinf/c$. 
This convective time interval corresponds to a Strouhal number ($St=fc/\Uinf$, with $f$ the reciprocal of the dimensional time interval) of \numrange{0.20}{0.25} (\cref{fig:forcepanel}). 

Subtle variations in the timing of the post-stall vortex shedding can lead to cycle-to-cycle variations during full stall. 
The most prominent factors contributing to these variations include boundary layer and shear layer instabilities, the three-dimensionality of the stall cell, fluid-structure interactions leading to vibrations of the airfoil, and free-stream turbulence or other perturbations in the free-stream velocity \citep{Harms2018, Snortland2023, damiola.2024a}.
The degree of cycle-to-cycle variations during full stall is indicated by the shaded areas in the lift evolution in \cref{fig:forcepanel}b, which shows the range between the minimum and maximum envelopes for the recorded cycles. 
The aerodynamic load fluctuations can introduce structural vibrations and lead to premature fatigue failure of wings and blades. 
Despite their significant impact, post-stall load variations have not always been treated with the proper level of diligence, as they are often concealed by phase averaging (\cref{fig:forcepanel}b). 

The cycle-to-cycle variations are not only prominent during the fully stalled stage, but they are still present when the lift coefficient starts to recover to its quasi-static value (between $t/T\approx 0.8$ and $t/T\approx 0.9$ in \cref{fig:forcepanel}b).
Shortly before the lift coefficient recovers, the pressure coefficient near the leading edge becomes negative again, indicating leading-edge suction recovery (\cref{fig:forcepanel}).
The suction recovery onset coincides with the global minimum of the lift coefficient, after which the lift coefficient gradually recovers to its quasi-static value (\cref{fig:forcepanel}).
The recovery of the leading-edge suction and lift coefficient are delayed to angles of attack well below the static stall angle of attack, which leads to the important dynamic stall lift hysteresis \citep{Williams2015,Williams2017}. 
We consider here the stall reattachment process to cover everything that happens between the moment when the geometric angle of attack falls below the critical stall angle $\kindex{\alpha}{ss}$ to the moment when the lift coefficient recovers to its quasi-static value given by \cref{eqn:qslift}.

As cycle-to-cycle variations make phase-averaged values inadequate to represent the post-stall and reattachment dynamics, we will continue analysing instantaneous time-resolved flow fields and aerodynamic loads to identify different flow regimes during dynamic stall reattachment. 
Our main goal will be to describe the different stages in the reattachment process, identify what triggers flow reattachment, extract critical parameters related to the onset of stall reattachment, and quantify the range of characteristic time-scales governing the reattachment process. 

\subsection{Flow development during dynamic stall reattachment}
\begin{figure}
\includegraphics[scale=1]{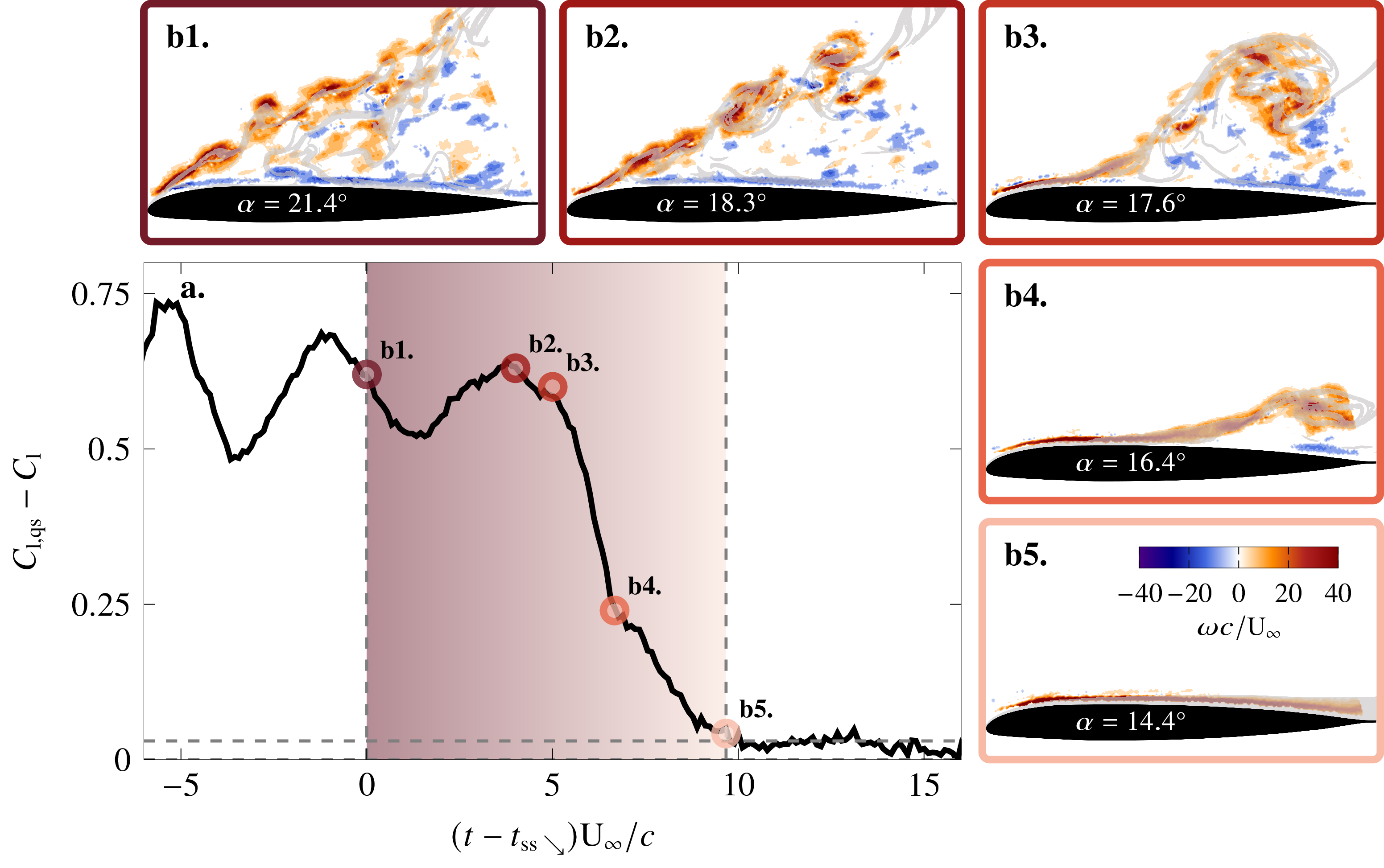}
\caption{(a) Temporal evolution of the lift deficit due to stall $(\Clqs - \kindex{C}{l})$ and selected snapshots of the vorticity and nFTLE fields during dynamic stall reattachment for the selected pitching cycle in \cref{fig:forcepanel}.
Snapshots (b1)-(b5) correspond to the marked instants b1-b5 on the lift deficit, ranging from the angle of attack dropping below the static stall angle (b1) to the point where the lift deficit converges to zero (b5). 
The range from (b1) to (b5) is highlighted by the shaded region and marks the entire dynamic stall reattachment process.}
\label{fig:flowpanel}
\end{figure}

Dynamic stall reattachment is a gradual process that evolves over time, similar to flow separation.
We describe this process using the temporal evolution of the lift deficit and selected flow-field snapshots (\cref{fig:flowpanel}).
The lift deficit is calculated as the difference between the quasi-static lift prediction in the absence of flow separation according to \cref{eqn:qslift} and the measured lift: $\kindex{\Delta C}{l,stall}=(\Clqs-\kindex{C}{l})$. 
The curve in \cref{fig:flowpanel}a shows the lift deficit during dynamic stall reattachment for the selected representative stall cycle presented earlier.
Snapshots (b1)-(b5) in \cref{fig:flowpanel} show the development of the flow above the airfoil during dynamic stall reattachment for the selected cycle. 
These snapshots of the instantaneous vorticity fields are captured between the moment the angle of attack drops below the critical stall angle ($\alpha=\ang{21.4}$) and the moment the lift deficit converges to zero.
This interval is highlighted in the lift deficit panel in \cref{fig:flowpanel}a and covers the entire dynamic stall reattachment process.
The flow fields are rotated in the airfoil frame of reference. 
Prominent ridges in the nFTLE fields are overlaid on the vorticity field to highlight the location and shape of the shear layer. 
The timing of the individual snapshots is indicated by the markers in \cref{fig:flowpanel}a. 

At the critical static stall angle during the pitch-down motion, the flow is fully separated and is characterised by a large separation region with vortices of both positive and negative vorticity (\cref{fig:flowpanel}b1).
A shear layer with strong positive shear layer vortices marks the boundary between the large separated flow region and the free-stream. 
At an angle of attack well below the stall angle, we see the first sign of flow reattachment in the flow field near the leading edge where the shear layer bends towards the surface of the airfoil and starts to reattach (\cref{fig:flowpanel}b2).
As the angle of attack continues to decrease, the shear layer progressively reattaches to the suction side from the leading to the trailing edge (\cref{fig:flowpanel}b3-b5), which allows the lift coefficient to increase and recover towards its quasi-static value. 
The evolution of the shear layer during this part of the cycle visually resembles the propagation of a whip wave similar to the ramp-down wave observed by \citet{Green:1995bq} in smoke flow visualisations. 
The measured lift coefficient finally catches up with its quasi-static prediction ($\Clqs-\kindex{C}{l}\to 0$) shortly after the shear layer reattachment wave has reached the trailing edge and the flow is fully reattached to the airfoil surface at $\alpha = \ang{14.4}$ (\cref{fig:flowpanel}b5).


\subsection{Dynamic stall reattachment stages}	
\begin{figure}
\includegraphics[scale=1]{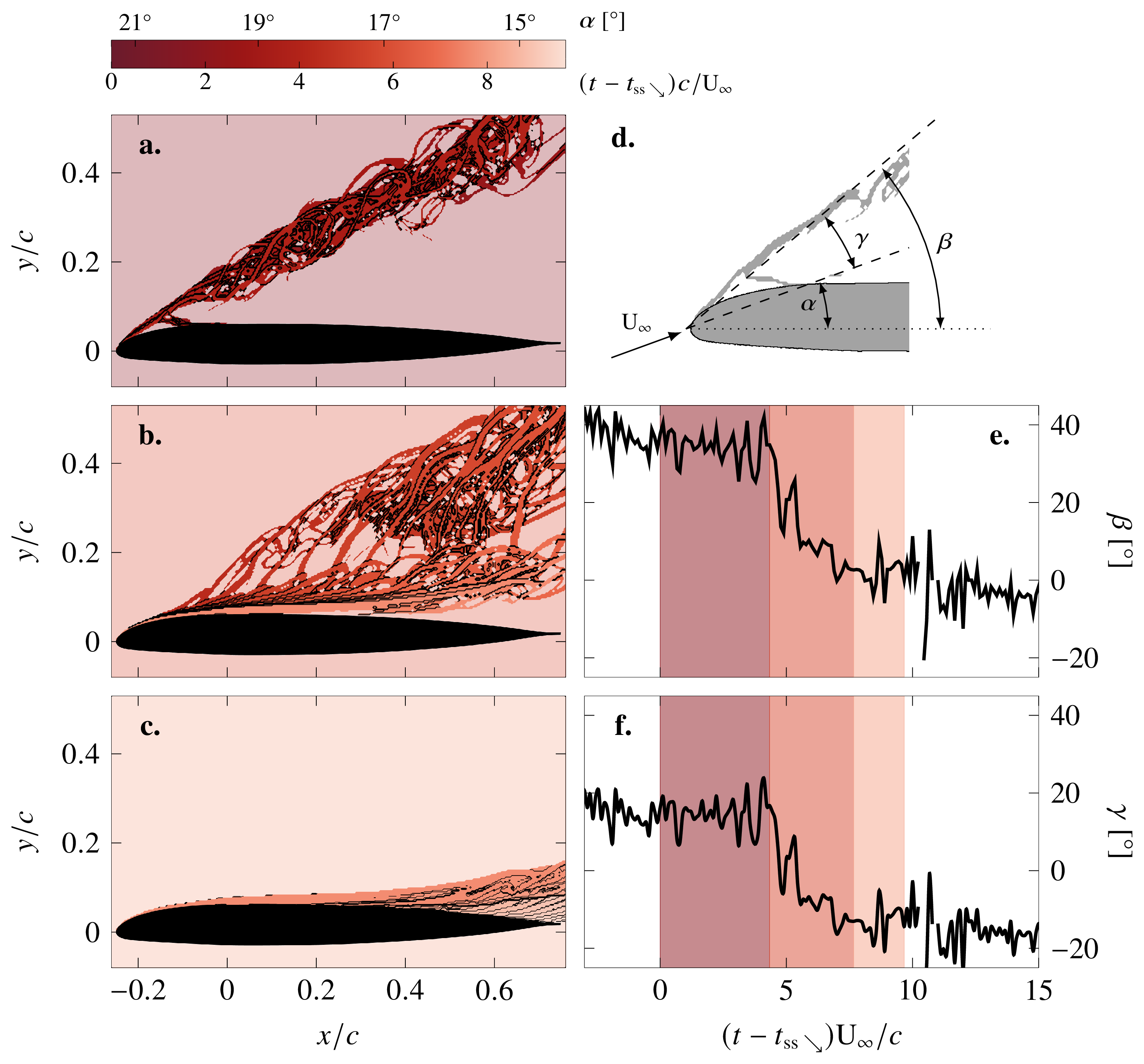}
\caption{(a-c) All nFTLE ridges extracted during the reattachment process, grouped into three time intervals.
Ridges are coloured based on the timing and angle of attack of the instantaneous snapshots from which they were extracted.
(d) Schematic illustration of the definitions of the angle of attack $\alpha$, ridge angle relative to the chord $\beta$, and ridge angle relative to the incoming flow $\gamma = (\beta - \alpha$).
(e) Temporal evolution of the shear layer angle relative to the chord $\beta$, for the selected cycle.
(f) Temporal evolution of the shear layer angle relative to the incoming flow direction ($\gamma$).
The shaded areas in (e) and (f) correspond to the duration of the three intervals indicated in (a)-(c) for the selected pitch cycle.}
\label{fig:ftleridges}
\end{figure}

The shear layer dynamics during the stall reattachment stages are summarised in \cref{fig:ftleridges}.
Here, the shear layer is represented by the most prominent ridge in the negative finite-time Lyapunov exponent (nFTLE) field.
The nFTLE ridges are colour-coded by time to highlight their temporal behaviour.
Based on the observed differences in the shear layer shape and dynamics during flow reattachment, we distinguish three stages in the flow reattachment process and combine the ridges observed during the three stages in separate plots (\cref{fig:ftleridges}a-c). 
The evolution of the nFTLE ridges is further quantified by the local angle near the leading edge between the most upstream part of the ridge that is detached from the airfoil surface and the airfoil chord, $\beta$ (\cref{fig:ftleridges}d,e), and the angle between the ridge and the incoming flow, $\gamma = \beta-\alpha$ (\cref{fig:ftleridges}d,f). 
The virtual intersection of the ridge and the chord moves downstream along the chord when the reattachment progresses. 
The shaded areas in \cref{fig:ftleridges}e,f correspond to the three reattachment stages. 

When the angle of attack first drops below the critical stall angle, the flow remains fully separated and the separated region is bound by a shear layer with small oscillations around a virtual straight line. 
Due to the unsteadiness of the pitching motion, there is a delay in the reaction of the flow to start reattaching similar to the reaction delay observed for stall onset.  
We call this stage the \textit{reaction delay} stage.

During the reaction delay stage, the flow remains fully separated, but the shear layer angle with respect to the airfoil chord ($\beta$) decreases, bringing the shear layer closer to the airfoil surface.
The increased proximity of the shear layer might support the initiation of flow reattachment and the transition to the second stage.  
The shear layer angle with respect to the airfoil chord decrease at the same rate as the angle of attack, such that the shear layer angle with respect to the incoming flow ($\gamma= \beta - \alpha$) remains at the approximately constant value of $\ang{20}$ throughout the entire stage. 

The reaction delay stage ends and the second stage begins when the actual stall recovery is initiated.
The initiation of stall recovery in the flow field is identified by the local change in curvature of the shear layer near the leading edge (\cref{fig:flowpanel}b).
This moment corresponds to a local maximum of the lift deficit due to stall (point b in \cref{fig:flowpanel}).
During this second stage, the front part of the shear layer progressively bends down, and its downstream part pushes the fully separated flow towards the trailing edge, visually resembling a whip wave convecting downstream.
During this part of the process, the flow field and the forces undergo the most important changes. 
We call this stage the \textit{wave propagation} stage. 

During the wave propagation stage, the shear layer gradually reattaches to the suction side and pushes the remnants of separated flow downstream.
The shear layer angle with respect to the incoming flow ($\gamma = \beta - \alpha$) decreases as the wave propagates over the airfoil surface. 
The decrease in the shear layer angle with respect to the airfoil chord ($\beta$) is significantly larger than the decrease in the angle of attack ($\alpha$) during this stage (\cref{fig:ftleridges}e,f).
This indicates that the wave propagation is driven by the local flow velocity and not by the pitching kinematics. 

When the shear layer reattachment wave reaches the airfoil trailing edge, marking the end of the wave propagation stage, the lift coefficient has not yet fully recovered.
The relaxation of the aerodynamic forces takes place during the final stages, and we refer to this state as the \textit{relaxation} state. 
Factors affecting this relaxation may be attributed to events within the boundary layer or to wake-induced effects. 
Additional experiments beyond the scope of this paper would be required to investigate these effects in detail. 

\subsection{Surface footprints of dynamic stall reattachment}

\begin{figure}
\centering
\includegraphics[scale=1]{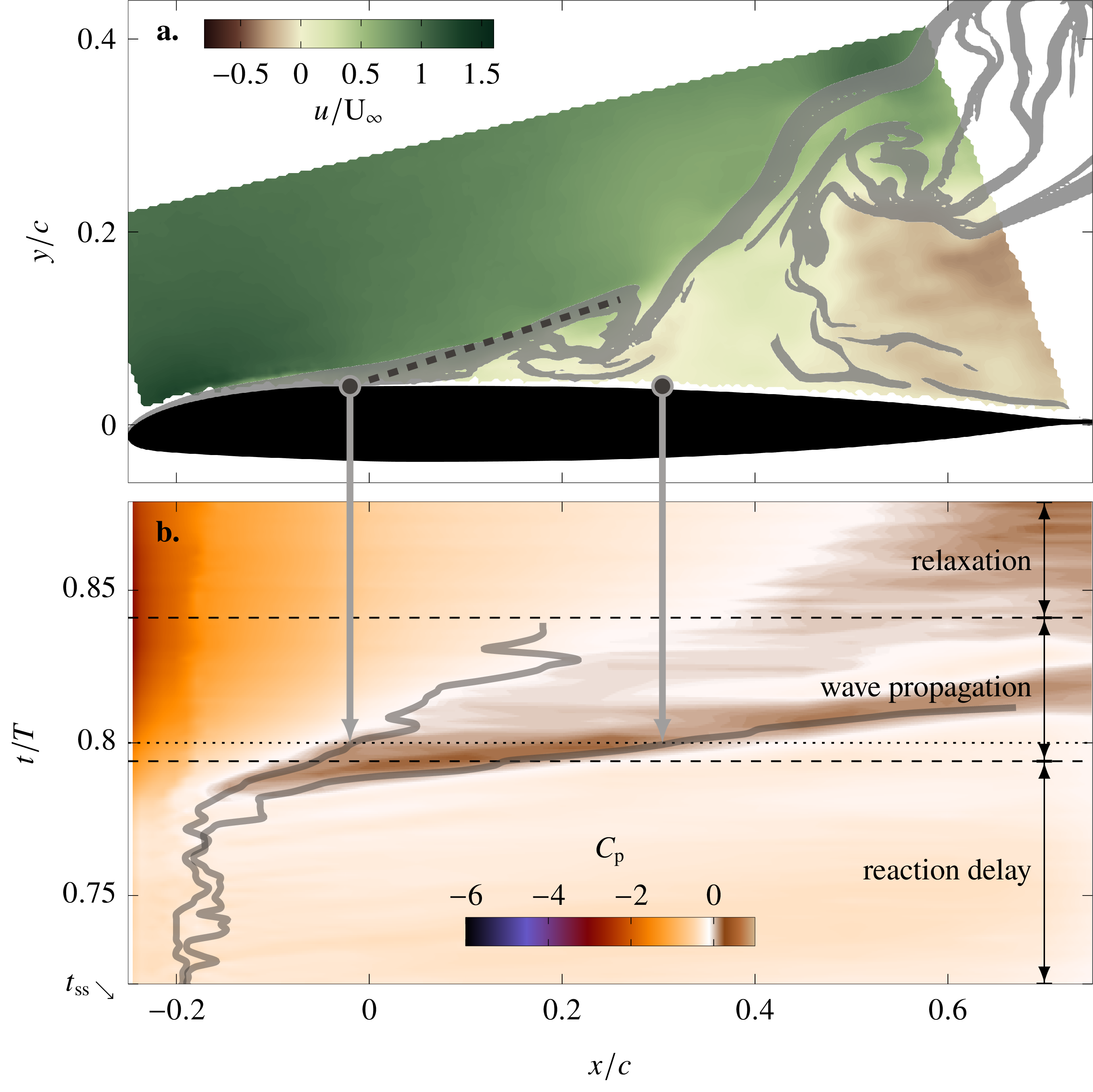}
\caption{(a) Example snapshot of the nFTLE ridge and horizontal velocity component ($u/\Uinf$) during the reattachment process ($\alpha=\ang{17.3}$, $t/T=0.81$). 
The transition points identified by the surface velocity reversal point ($\kindex{u}{surf}=0$) and the nFTLE ridge intersection are marked on the airfoil.
(b) Temporal evolution of the transition points overlaid on the surface pressure field during the pitch-down part of the cycle.}
\label{fig:points_CP}
\end{figure}

The nFTLE ridge that represents the shear layer is an attractor or separation line and forms the boundary between the separated flow above the airfoil and the incoming or attached flow. 
Ideally, the separation point would be the intersection of the separation line with the airfoil surface. 
Due to the no-slip boundary condition at the wall, FTLE ridges do not make direct contact with the surface \citep{Serra2018, Klose2020}.
Lagrangian flow separation begins with the upwelling of fluid material from the wall, creating sharp spikes in material lines initially parallel to the surface \citep{Serra2018}. 
The theoretical centerpiece of these spikes is called the Lagrangian backbone of separation. 
The intersection of the backbone of separation with the wall is the spiking point, which marks where separation initiates in the Lagrangian frame. 
This spiking point can be identified from severe curvature changes in advected material lines or from high-order derivatives of the wall-normal velocity \citep{Serra2018, Klose2020}. 
This approach requires high spatial resolution to extract the high-order derivatives. 
The spiking is typically located upstream of the Prandtl separation point, which refers to the location of zero skin friction. 

In the present work, we adopt a simplified approach to approximate the Lagrangian separation point from the FTLE field. 
We extrapolate the linear section of the nFTLE ridge close to the airfoil surface to find its intersection with the wall (\cref{fig:points_CP}a). 
This approximation estimates where the nFTLE ridge approaches the surface, and is used here as a proxy for the Lagrangian spiking point.

As an estimate of the Prandtl separation point, which we refer to as the velocity-based separation point, we use the most upstream flow reversal point in the near-surface velocity $\kindex{u}{surf}$ (\cref{fig:points_CP}).
We consider the horizontal velocity component in the airfoil frame of reference in the second-closest grid point to the airfoil surface as the near-surface velocity. 
Due to the geometry of the airfoil, the difference between the horizontal velocity component and the tangential component is minor along most of the chord length, but the former is easier to extract. 
The most upstream chordwise location where the near surface velocity changes sign from positive to negative ($\kindex{u}{surf}=0$) is taken as the velocity-based separation point.

The presented snapshot represents the wave-propagation state during flow reattachment (\cref{fig:points_CP}a).
The location of the nFTLE root in this state is found upstream of the surface velocity reversal point, which is expected as the nFTLE ridge marks the outer boundary of the separated flow region, which is characterised by fluid with a lower kinetic energy that is not necessarily reversed when analysed in an instantaneous manner \citep{Serra2018}.  
The spatio-temporal evolution of the nFTLE and surface velocity-based separation points during the full reattachment process are presented in \cref{fig:points_CP}b on top of the surface pressure distribution for the example pitching cycle presented before. 
As a direct consequence of the extraction procedure, the evolution of nFTLE-based separation point is subject to larger fluctuations than the surface velocity-based separation point. 
The more the nFTLE ridge is aligned with the airfoil, the harder it gets to extract a meaningful nFTLE-based separation point. 
Therefore, we only show the surface trace of the nFTLE root in \cref{fig:points_CP}b until the end of the wave propagation stage. 

During the reaction delay stage, the nFTLE-based and velocity-based separation points are close to the leading edge, at approximately $x/c=-0.2$, which is $0.05c$ downstream of the leading edge.
Both separation points start to move towards the trailing edge at the start of the wave propagation stage, but they follow different trajectories that leave distinct features in the surface pressure distribution. 
The trajectory of the surface velocity-based separation point follows the maximum pressure ridge in the surface pressure distribution (\cref{fig:points_CP}b).
The surface velocity-based separation point moves with an approximately constant speed of $0.41\Uinf$ and reaches the trailing edge well before the shear layer wave does. 
The motion of the shear layer wave is slower as it has to push the entire separated flow region downstream. 

The trajectory of the nFTLE-based separation point that marks the initial progression of the shear layer wave aligns closely with the zero-contour of the pressure coefficient along the chord (\cref{fig:points_CP}b). 
Upstream of the nFTLE-based separation point, the flow reattaches, and we expect suction recovery. 
The trace of the nFTLE-based separation point in \cref{fig:points_CP}b does not reach the trailing edge at the end of the wave propagation stage because we cannot reliably estimate the nFTLE-based separation point when the ridge angle with respect to the airfoil surface becomes small. 

What exactly triggers the shear layer curvature and initiates the stall recovery process is not yet clear. 
As the propagation of the extracted separation points and the footprints in the pressure distribution all move from the leading towards the trailing edge, we focus next on the leading-edge suction parameter to identify the necessary and sufficient conditions that trigger the onset of stall recovery. 

\begin{figure}
\centering
\includegraphics[scale=1]{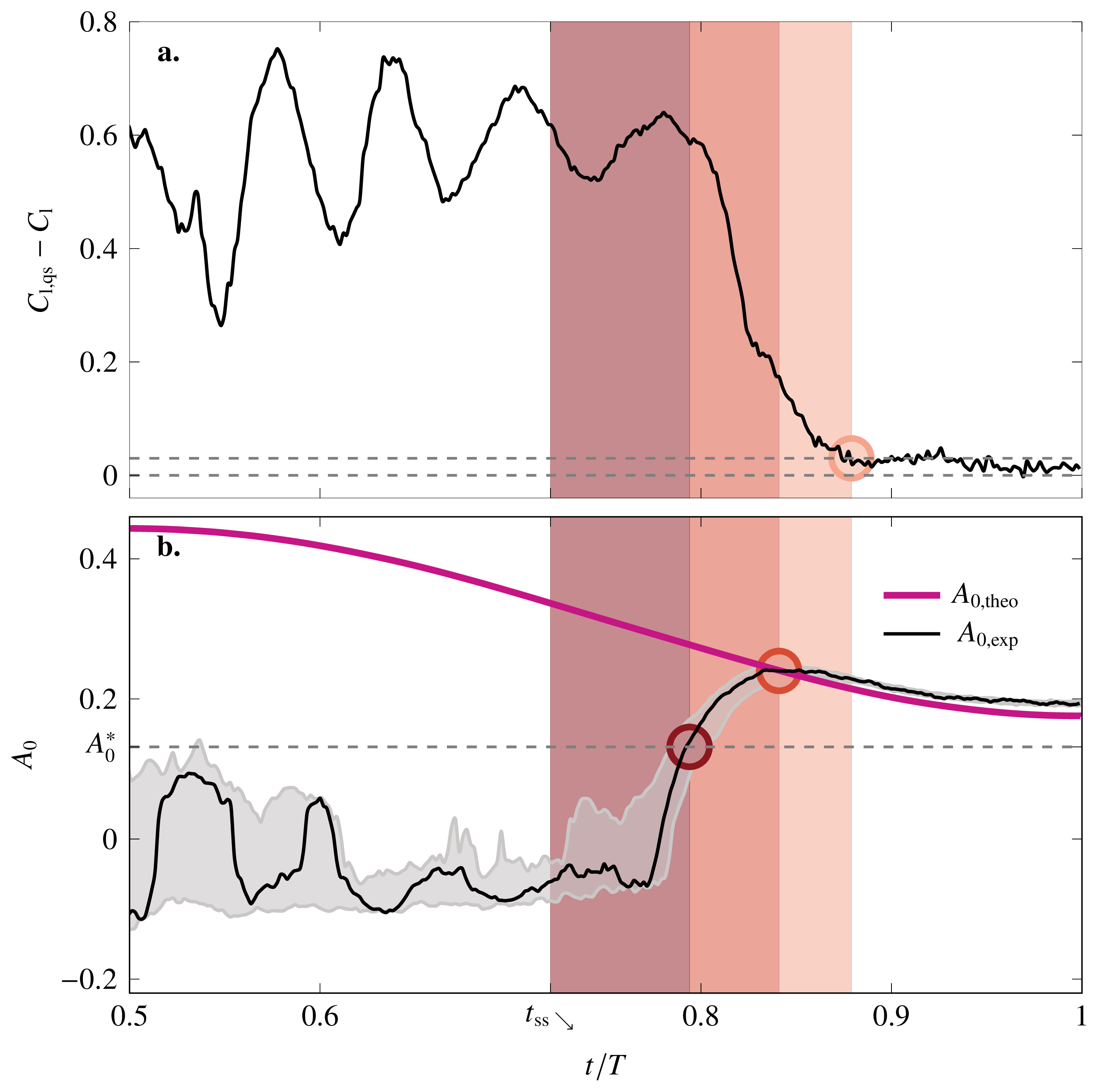}
\caption{(a) Temporal evolution of the lift deficit due to stall ($\Clqs- \kindex{C}{l} \to 0$) and (b) the leading-edge suction parameter for the pitching motion presented in \cref{fig:forcepanel}.
The values of the representative cycle are shown in black, and the gray shaded area shows the range of cycle-to-cycle variations across the all recorded cycles.
The theoretical leading-edge suction parameter ($\kindex{A}{0,theo}$) is shown for comparison to the experimental values in (b).
The colour-shaded regions indicate the three stages of the dynamic stall reattachment for the selected cycle. 
The colour-shaded regions are the same as in \cref{fig:ftleridges}. 
The transition points between the states are indicated with markers. The value of the leading-edge suction parameter at the onset of wave-propagation state ($\kindex{A}{0}^*$) and is indicated by the horizontal dashed line in (b).     
}
\label{fig:A0panel}
\end{figure}

\subsection{Critical leading-edge suction parameter at reattachment onset}\label{sec:criticalA0}

\Cref{fig:A0panel} summarises the temporal evolutions of the leading-edge suction parameter $\kindex{A}{0}$, its theoretical prediction for attached flow $\kindex{A}{0,theo}$, and the lift deficit due to stall $\kindex{\Delta C}{l,stall}=\Clqs-\kindex{C}{l}$ for the representative cycle of the example pitching motion presented before. 
The leading-edge suction parameter is calculated based on the pressure data from the front \qty{10}{\percent} of the chord following \citet{Deparday2019} and \citet{He.2020}.
The theoretical prediction of the leading-edge suction parameter for a sinusoidally pitching airfoil and attached flow is also calculated following \citet{Deparday2019}: 
\begin{equation}
\kindex{A}{0,theo}(t) = \sin\alpha(t) + \dot{\alpha}(t)\frac{c}{4\Uinf} - \kindex{K}{$\eta$} \cos\alpha(t),
\label{Eq:A0theory}
\end{equation}
where $\dot{\alpha}(t)$ is the pitch rate and $\kindex{K}{$\eta$}$ represents the effect of the airfoil camber.
The chord-normal coordinate of the camber line along the chord is $\eta(x)$, such that:
\begin{equation}
\kindex{K}{$\eta$} = \frac{1}{\pi} \int\limits_{0}^{\pi} \dv{\eta(\theta)}{x} \dd{\theta}. 
\end{equation}
The evolution of the experimetally determined leading-edge suction parameter for the representative example cycle is shown in black, and the range of cycle-to-cycle variations are represented by the gray shaded area between the minimum and maximum envelopes extracted across all measured cycles.
The shaded areas in colour are the same intervals identified in \cref{fig:ftleridges}.
These areas correspond to the three stages of the reattachment process that starts when the angle of attack drops below the static stall angle and is considered finished once the lift coefficient recovers to its unstalled values. 
The latter is equivalent to the lift deficit due to stall dropping below \qty{3}{\percent} as indicated by the marker in \cref{fig:A0panel}a. 

\begin{figure}
\centering
\includegraphics[scale=1]{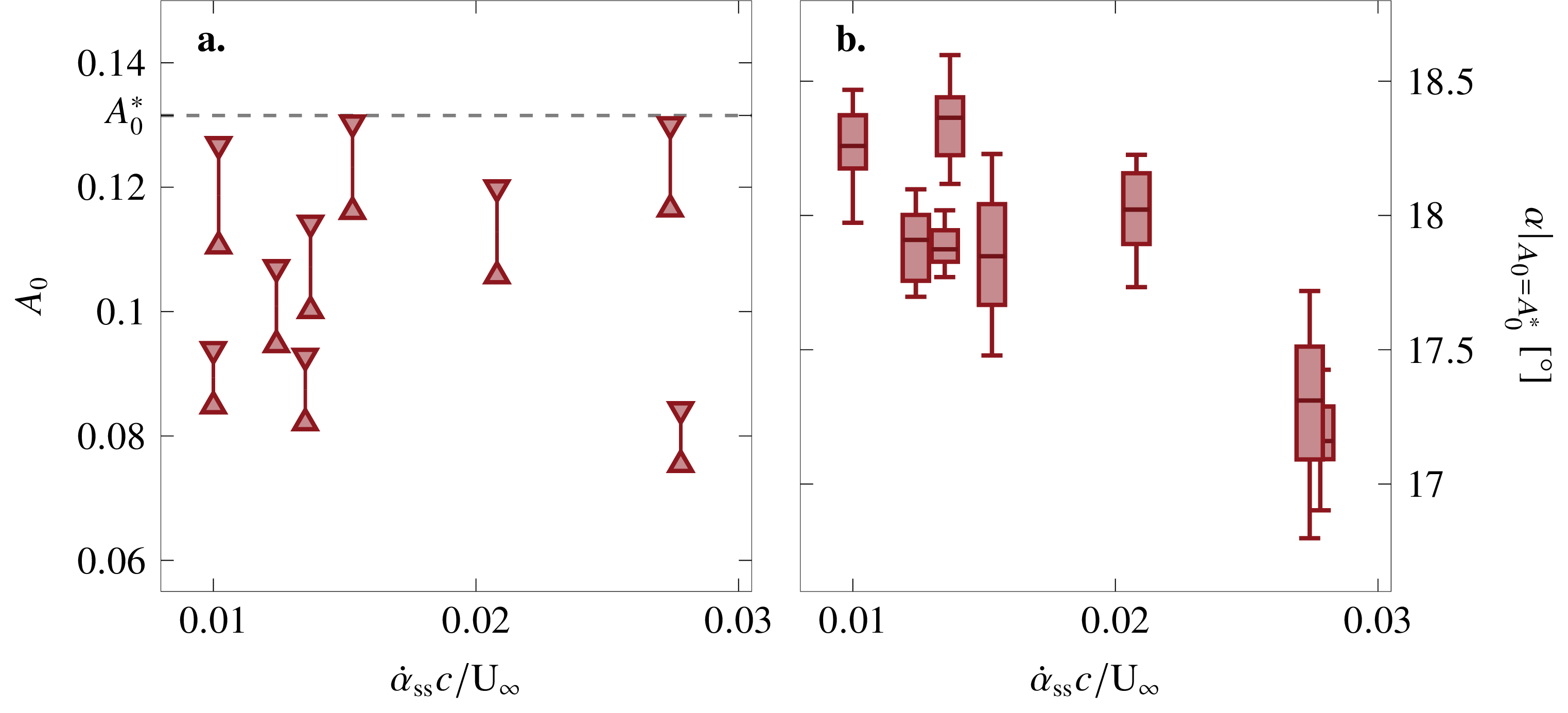}
\caption{(a) Range of values of the leading-edge suction parameter that work as a critical threshold for stall recovery in \qty{95}{\percent} (bottom marker) and \qty{98}{\percent} (top marker) of the cycles as a function of the non-dimensional pitch rate at static stall. 
The critical value $\kindex{A}{0}^*$ corresponds to the upper bound of this range indicated by the dashed line.
(b) Angle of attack distribution at $\kindex{A}{0}^*$ as a function of the non-dimensional pitch rate.}
\label{fig:A0_critical}
\end{figure}

The onset of the wave propagation stage and the true start of the stall recovery is delayed with respect to the moment when the angle of attack drops below the critical stall angle. 
To initiate stall recovery, it is necessary that the angle of attack is below the critical stall angle, but this is clearly not a sufficient condition for dynamic stall recovery. 
For various deep dynamic stall cases analysed here, the angle of attack at the onset of stall recovery varies with the unsteadiness of the pitching motion expressed by $\adottext$ (\cref{fig:A0_critical}b). 

In analogy with the stall onset criterion based on the leading-edge suction parameter introduced by \citet{Ramesh2013}, we hypothesise that the onset of stall recovery can only occur when a critical leading edge suction is exceeded. 
The stall onset criterion is based on the idea that an airfoil can only support a maximum amount of leading edge suction. 
If this critical leading edge suction is exceeded during the pitch-up motion, vorticity in the shear layer has to be released into the flow leading to stall onset.
Inversely, we hypothesise that a minimum amount of leading edge suction is required to pull the vorticity in the shear layer back to the airfoil surface and initiate the onset of recovery. 

Here, we check whether we can find a critical value of the leading-edge suction parameter that is a necessary and sufficient condition for stall recovery. 
The candidate critical leading-edge suction parameter value is the maximum value of the leading-edge suction parameter observed during the reaction delay and fully separated flow stages, as indicated by the horizontal dashed line in \cref{fig:A0panel}b.
Due to existing cycle-to-cycle variations and measurement noise, we determined the critical leading-edge suction parameter through statistical analysis of the pressure data collected from various oscillation cycles. 
To determine this threshold, we tested a range of candidate values and evaluated their effectiveness in initiating recovery across individual cycles in each dataset. 
Recovery onset was confirmed when the leading-edge suction parameter consistently stayed above the threshold value after initially exceeding it. 
If the parameter dropped significantly below the threshold after initially exceeding it, the flow returned to a separated state, and the threshold was insufficient to trigger recovery.
A detailed explanation of this procedure is given in ~\cref{appendix:A0statistical}.

Following the approach in~\cref{appendix:A0statistical}, the critical leading-edge suction parameter was obtained for different pitch rate cases and presented in \cref{fig:A0_critical}a.
The bottom and top markers in \cref{fig:A0_critical}a correspond to values of the leading-edge suction parameter that work as a critical threshold for stall recovery in respectively \qty{95}{\percent} and \qty{98}{\percent} of the cycles for a given pitching motion.
This range of values gives us a measure for the uncertainty range in determining the critical value of the leading suction parameter for stall recovery. 
The critical leading-edge suction parameter values range between \numrange[range-phrase = {\text{\ and\ }}]{0.07}{0.13} with no clear dependence on the pitch rate. 
The unsteadiness of the pitching motion is characterised by the effective non-dimensional pitch rate introduced in \citet{Mulleners2012}, which is defined as $\adottext$ with $\kindex{\dot{\alpha}}{ss}$ the pitch rate at the moment the static stall angle is exceeded.
The maximum threshold value across all pitch rates, $\kindex{A}{0}^*=0.13$, was selected as the critical leading-edge suction parameter for the conditions of our measurements (dashed horizontal line in \cref{fig:A0panel}b).
The maximum value is chosen as the critical value to ensure the onset of recovery for all cases operating at this Reynolds number for the tested airfoil.
The angle of attack at the moment when the leading-edge suction parameter exceeds $\kindex{A}{0}^*$ during the pitch-down motion is not independent of the pitch rate, but decreases with increasing effective non-dimensional pitch rate.

The critical leading-edge suction parameter is not a universal value, but rather a critical threshold that serves as a necessary and sufficient condition for stall recovery. 
The value of the critical leading-edge suction parameter may vary with Reynolds number and airfoil configuration and should be validated for specific flow conditions and airfoil configurations. 
The values of the experimentally extracted leading-edge suction parameter also depend on the number and location of the pressure sensors included in the calculation \citep{Deparday.2022}. 
Here, we determined the leading-edge suction parameter based on the pressure signals from the sensors located in the first \qty{10}{\percent} of the airfoil similar to what was done in previous work by \cite{Deparday2019}. 
For the airfoil geometry considered here, the critical leading-edge suction parameter reaches a plateau for an integration range that includes at least the front \qty{10}{\percent} (\cref{appendix:LESPintegral}).
For the subsequent analysis of the characteristic time-scales of the reattachment process, we use $\kindex{A}{0}^*=0.13$ as the value critical leading-edge suction parameter.

The comparison between the experimental and theoretical values of the leading-edge suction parameter suggests another characteristic milestone in the stall reattachment process that we can systematically calculate.
The end of the wave-propagation stage was identified previously based on the analysis of the shear layer ridge angle evolution, which requires time-resolved flow field data and is not easily accessible. 
The moment the shear layer aligns with the airfoil surface and its relative angle to the flow, $\gamma$, reaches zero corresponds to the intersection of the experimental and theoretical curves of the leading-edge suction parameter (\cref{fig:A0panel}b).
Based on this observation, we suggest to determine the end of the wave-propagation stage when the leading-edge suction parameter recovers to the theoretical value, which is a more easily accessible pressure-based indicator and eliminates the need for flow field data.

\subsection{Characteristic time-scales of the reattachment process}
Now that we have determined robust pressure-based indicators of the transition points between the different stall reattachment stages, we can extract the characteristic time scales governing the process that can be used in the future to refine dynamic stall models and to develop effective control strategies. 
Understanding the time scales of the recovery process is particularly interesting when using semi-empirical dynamic stall models like the Beddoes-Leishman and Goman-Khrabrov models.
The key input parameters of these models often relate to characteristic time scales of the prominent stages in the flow and forces development and their dependence on the unsteadiness of the pitching kinematics.

\begin{figure}
\centering
\includegraphics[scale=1]{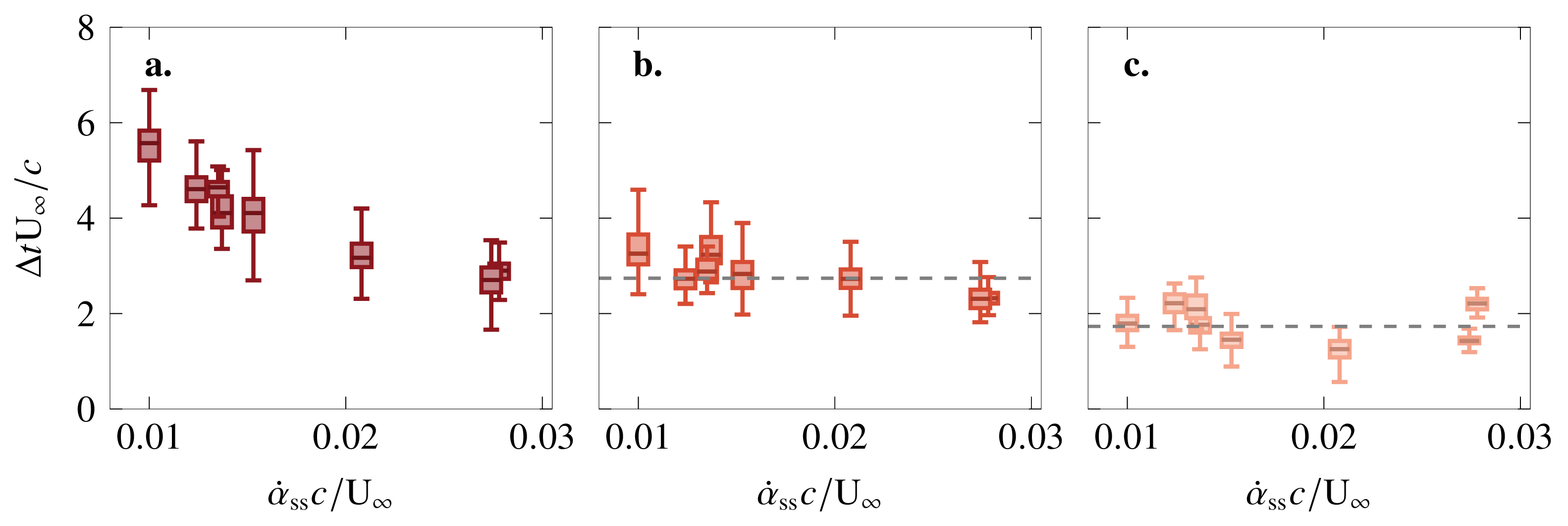}
\caption{Distribution of the time delays corresponding to the
(a) reaction delay stage,
(b) wave propagation stage, and
(c) relaxation stage as a function of the effective pitch rate. Horizontal dashed lines in (b) and (c) show the average timing for these states across all pitch rates. The timescales are non-dimensionalised using the convective time $\Uinf/c$.}
\label{fig:boxplots}
\end{figure}

The characteristic time scales corresponding to the different reattachment stages are extracted for each recorded pitching cycle for different pitching kinematics. 
The distribution of the non-dimensional time scales ($\Delta t \Uinf/c$) are presented in the form of boxplots in \cref{fig:boxplots} as a function of the effective non-dimensional pitch rate.
The horizontal streak in each box represents the median value, and the edges of the box indicate the interquartile range, capturing the middle \qty{50}{\percent} of the data.
Whiskers extend to the minimum and maximum observed timings, highlighting the full range of the extracted time delays per case.

The reaction delay stage covers the interval between the moment the angle of attack falls below the critical stall angle and the onset of wave propagation marked by the moment the critical leading-edge suction parameter value $\kindex{A}{0}^*$ is exceeded.
The reaction delay decreases with increasing effective pitch rate (\cref{fig:boxplots}a). 
When the airfoil pitches down it increases the local effective velocity of the leading edge. 
Higher pitch rates lead to increased effective leading-edge velocities which are associated with increased leading-edge suction values.
The critical leading-edge suction value will thus be attained earlier when the pitch rate is increased. 
In \cref{fig:boxplots}a, we calculated the reaction delay between the moment the geometric angle of attack falls below the critical stall angle and the moment the critical leading-edge suction is attained. 
Due to the unsteady pitch down motion, the effective angle of attack is lower than the geometric angle of attack (\cref{appendix:alphaeff}). 
If we consider the start of the reaction delay as the moment the effective angle of attack falls below the critical stall angle, the reaction delays are slightly increased for all pitch rates, but the overall decrease in reaction delay with increasing pitch rate remains unaffected (\cref{fig:alphaeff}).

The wave propagation stage spans the period between the moment the critical leading-edge suction value is exceeded and the moment the leading-edge suction parameter recovers to its theoretical value.
The relaxation stage begins when the leading-edge suction parameter reaches its theoretical value and ends when the experimental lift coefficient converges to within \qty{3}{\percent} of its quasi-static value.
The delays associated to both the wave propagation and the relaxation stages are independent of the pitch rate (\cref{fig:boxplots}b,c). Across all pitch rates, the wave propagation state lasts an average of \num{2.7} convective times, and the relaxation state lasts \num{1.7} convective times, indicated by dashed lines in \cref{fig:boxplots}b,c.
The total reattachment delay is the sum of the delays of the three stages. 

We observe a spread in the measured time scales across the different pitching cycles.
The range of reaction delay for each pitch rate case ranges from approximately \numrange{1}{4} convective times.
This spread is not a measurement artefact but is mainly due to the inherent cycle-to-cycle variations, similar to those observed in the post-stall vortex shedding process.
The recovery onset is subject to cycle-to-cycle variations, and it is necessary to account for this inherent variability when modelling dynamic stall reattachment. 
Incorporating an uncertainty range into reattachment onset predictions will improve the modelling accuracy and robustness.

\section{Conclusion}

Flow reattachment is the final stage of the dynamic stall cycle, where the lift coefficient recovers to its quasi-static value after being lost due to flow separation.
We investigated the stall recovery process for a sinusoidally pitching airfoil by analysing the shear layer dynamics derived from flow field data and by examining significant changes in the lift coefficient and leading-edge suction parameter obtained from pressure field data.
We identified what triggers stall recovery, described the sequence of the events during the recovery process, and determined their characteristic time scales.

A fully separated flow does not immediately start to recover when the angle of attack falls below the critical stall angle but is delayed to lower angles of attack.
Only when the shear layer is curved towards the leading edge, it allows for the leading edge suction, and ultimately the lift, to start recovering. 
The progression of the suction recovery is preceded by a shear layer wave that propagates from the leading to the trailing edge and pushes reversed flow downstream. 
This shear layer wave is convected by the local flow velocity and basically cleans the airfoil from low-momentum fluid, paving the way for suction recovery. 

We identified a necessary and sufficient condition for the onset of stall recovery in the form of a critical value of the leading-edge suction parameter.
The critical value is found to be independent of the pitch rate of the kinematics.

The recovery process consists of three stages, which we refer to as the reaction delay, the wave propagation, and the relaxation stage. 	
The reaction delay stage covers the interval between the moment the angle of attack drops below the critical stall angle and the moment the critical leading-edge suction parameter is reached. 
The flow during the reaction delay stage remains fully separated. 
Once the leading-edge suction parameter reaches a critical value, the actual stall recovery is initiated by the downstream convection of the shear layer wave during the wave propagation stage. 
The shear-layer wave pushes the remnants of separated flow downstream, which is an essential precursor for actual reattachment during the relaxation stage. 
At the transition between the wave propagation and the relaxation stage, the leading-edge suction parameter returns to its theoretical values. 
The end of the stall recovery process is marked by the recovery of the lift coefficient to its quasi-static values. 

The duration of the reaction delay state decreases with increasing effective pitch rate and was the longest stage of the process for the experimental settings analysed here. 
The durations of the wave propagation and the relaxation stage are independent of the pitch rate. 
The reaction delay experiences cycle-to-cycle variations up to \num{4} convective times, which should be taking into account in the modelling of stall recovery.

This paper contributes to a better understanding of the recovery process by providing a phenomenological description of the subsequent stages of the dynamic stall reattachment process. 
The combined use of pressure and flow field data revealed how footprints in each data type are linked.
These results support the development of control strategies for flow recovery in unsteady aerodynamic systems. 
The identification of a critical leading-edge suction parameter and the characteristic time scales associated with the stall reattachment stages can be used in future work to improve semi-empirical dynamic stall models including stall reattachment.
 



\backsection[Declaration of interests]{The authors report no conflict of interest.}




\appendix
\section{Presentation of additional cycles}\label{appendix:morecycles}
\begin{figure}
\centering
\includegraphics[scale=1]{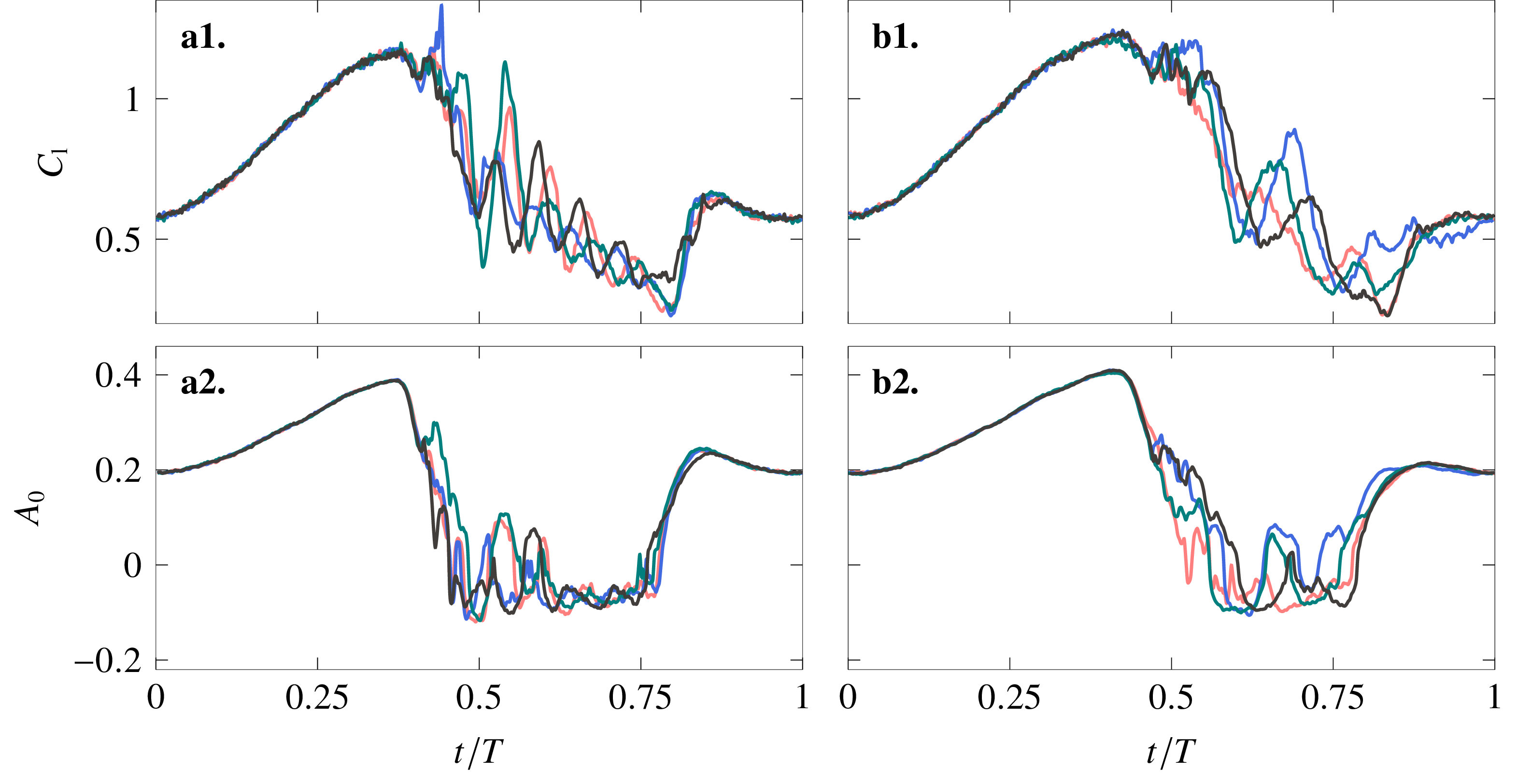}
\caption{(a1) Lift coefficient and (a2) leading edge suction parameter for four cycles of an oscillation with $\kindex{\alpha}{0}=\ang{20}$, $\kindex{\alpha}{1}=\ang{8}$, $k=\num{0.05}$, $\kindex{\dot{\alpha}}{ss}=\num{0.0135}$. 
The cycle in black is the sample case studied in the main text. 
(b1) Lift coefficient and (b2) leading edge suction parameter for four cycles of an oscillation with $\kindex{\alpha}{0}=\ang{20}$, $\kindex{\alpha}{1}=\ang{8}$, $k=\num{0.1}$, $\kindex{\dot{\alpha}}{ss}=\num{0.0274}$.}
\label{fig:more_cycles}
\end{figure}

The analysis presented in the main text is based on a single representative cycle.  
In addition to the minimum and maximum envelopes included in \cref{fig:forcepanel} and \cref{fig:A0panel}, we present in \cref{fig:more_cycles} data from additional cycles for the lift coefficient and the leading edge suction parameter for two pitching motions with different pitch rate.

Our analysis focuses on the second half of the pitch cycle ($t/T>0.5$), where the flow undergoes reattachment following an initially fully separated state. 
During this phase, both quantities exhibit fluctuations due to continuous vortex shedding in the fully separated state. 
These fluctuations are present across all cycles, but the precise timing and magnitude of individual peaks vary between cycles (\cref{fig:more_cycles}). 
In the fully separated regime, the selected cycle provides a qualitative representation of the flow behavior, though quantitative details differ between cycles. 
With the onset of flow recovery, the lift and leading edge suction parameter increase to higher values, and this increase is common across different cycles. 
Once the flow fully recovers, the differences between cycles disappear.

\section{Influence of unsteady pitching motion on the effective airfoil angle of attack}\label{appendix:alphaeff}
\begin{figure}
\includegraphics{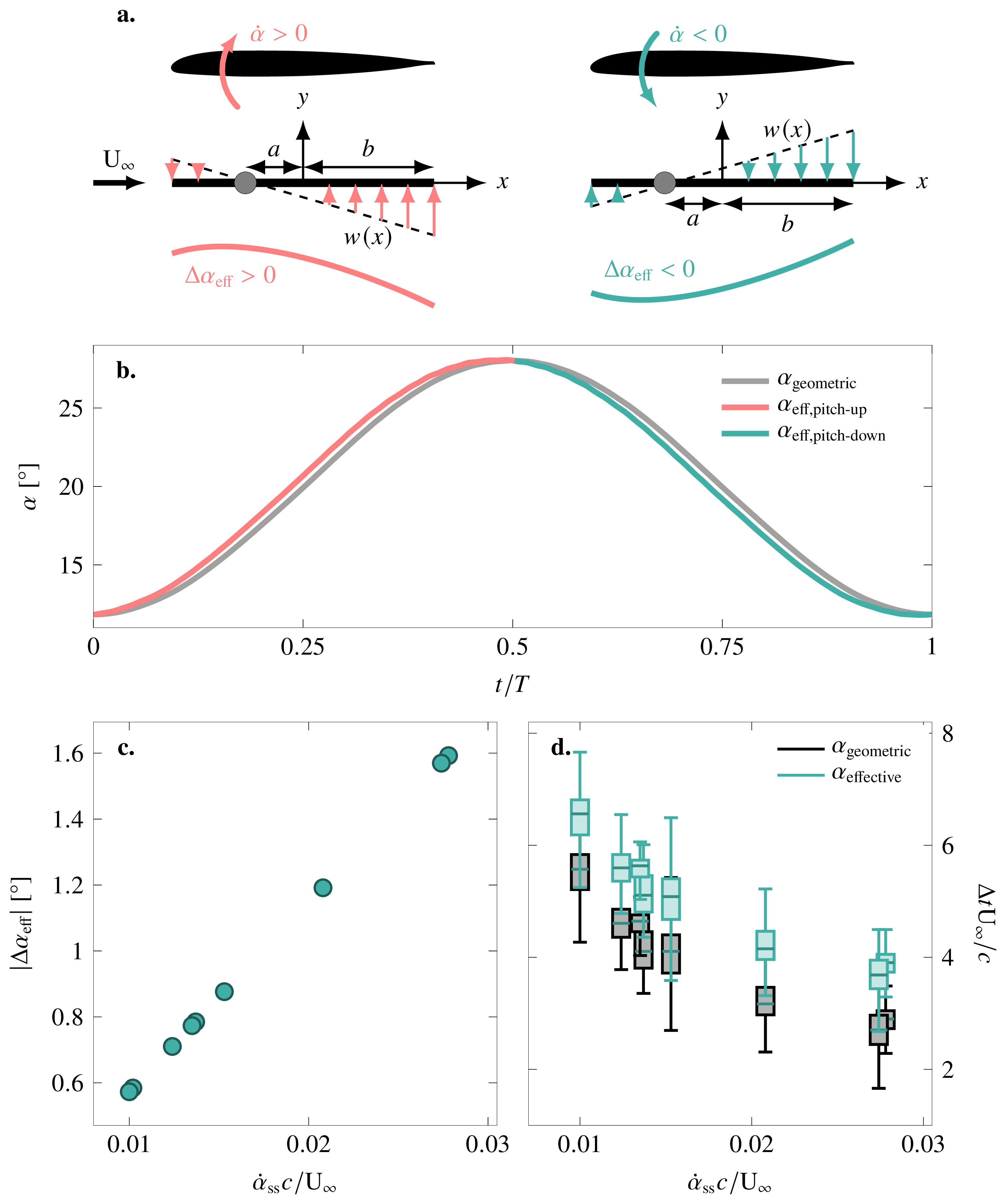}
\caption{
(a) Schematic illustration of the induced vertical velocity distribution ($w(x)$) and camber created by pitch-up and pitch-down motion. 
The illustrations are adapted from \cite{Leishman2006}. 
(b) Comparison of the evolutions of the effective and geometric angles of attack for a representative pitching motion ($\kindex{\alpha}{0}=\ang{20}$, $\kindex{\alpha}{1}=\ang{8}$, $k=\num{0.05}$, $\kindex{\dot{\alpha}}{ss}=\num{0.0135}$). 
(c) Magnitude of the angle of attack offset $|\Delta\alpha|$ as a function of the effective pitch rate.  
(d) Reattachment reaction delay determined using the effective or the geometric angle of attack as a function of the effective pitch rate. 
}
\label{fig:alphaeff}
\end{figure}

An unsteady pitching motion of an airfoil produces a chordwise distribution of induced vertical velocities along the chord ($w(x)$), that can be modelled as an effective change in airfoil camber (\cref{fig:alphaeff}a). 
The induced camber leads to an angle of attack variation, which can be calculated using thin airfoil theory \citep{Leishman2006, Brunton2012}. 

Consider an airfoil subjected to a pitching motion about a fixed pivot point with pitch rate $\dot{\alpha}$. 
The induced chord-normal velocity varies linearly along the airfoil chord.
In a coordinate system with the origin at mid-chord, and the pivot point location a distance $a$ from the origin, the induced chord-normal  velocity varies as: 
\begin{equation}
w(x) = -\dot{\alpha} (x - ab),
\end{equation}
with $b=c/2$ the semi-chord.

The first two Fourier coefficients are obtained by substituting this induced velocity into the standard definitions of classical thin airfoil theory and applying the coordinate transform $x=-b\cos\theta$:

\begin{align} \kindex{A}{0} &= \alpha - \frac{1}{\pi}\int\limits_0^{\pi} \frac{w(x)}{\Uinf} \dd\theta = \alpha - \frac{\dot{\alpha}c}{2\Uinf}a, \\ \kindex{A}{1} &= \frac{2}{\pi}\int\limits_0^{\pi} \frac{w(x)}{\Uinf} \cos\theta \dd\theta = \frac{\dot{\alpha}c}{2\Uinf}. \end{align}

The lift coefficient according to thin airfoil theory is given by:
\begin{align}
\kindex{C}{l} &= 2\pi\left(\kindex{A}{0}+\frac{\kindex{A}{1}}{2}\right)\nonumber\\
&=2\pi\left[ \alpha + \frac{\dot{\alpha}c}{4\Uinf}\left(1-2a\right)\right].\label{eq:tatCl}
\end{align}
The first term in \cref{eq:tatCl}, represents the lift coefficient under steady conditions.
The contribution to the lift coefficient due to the unsteady pitching motion can be seen as a correction to the angle of attack:
\begin{align}
\Delta\alpha = \kindex{\alpha}{eff} - \alpha,
\end{align}
with \kindex{\alpha}{eff} the effective angle of attack and
\begin{equation}
\Delta\alpha=\frac{\dot{\alpha}c}{4\Uinf}\left(1-2a\right).
\end{equation} 


The pitching axis of the airfoil considered in this paper is located at the quarter-chord ($a = -1/2$).
For this configuration, the effective angle of attack simplifies to
\begin{align}
\kindex{\alpha}{eff} = \alpha + \frac{\dot{\alpha}c}{2\Uinf}.
\end{align}
During a pitch-up motion ($\dot{\alpha}>0$), the induced positive camber increases the effective angle of attack $\kindex{\alpha}{eff}$ relative to the geometric angle of attack, as shown in the first half of the cycle in \cref{fig:alphaeff}b. 
During a pitch-down motion ($\dot{\alpha}<0$), the induced negative camber decreases $\kindex{\alpha}{eff}$, making it lower than the geometric angle of attack, as shown in the second half of the cycle in \cref{fig:alphaeff}b.

For the sinusoidal pitching motions considered in this paper, the maximum difference between the geometric and effective angles of attack occurs at the mean angle of attack of the pitching motion and depends on the pitch rate. 
The mean angles of attack for the motions considered here are $\kindex{\alpha}{0} \in \{\ang{18}, \ang{20}, \ang{22}\}$, which are close to the critical static stall angle of attack. 
For the motions considered here, the difference between the geometric and effective angles of attack $\Delta \alpha$ at the moment the geometric angle of attack equals the critical static stall angle, which we refer to as the effective angle of attack offset, increases with increasing effective pitch rate $\adottext$ (\cref{fig:alphaeff}c). 
The effective angle of attack is $\approx \ang{0.5}$ to $\ang{1.5}$ higher (when pitching up) or lower (when pitching down) than the geometric angle of attack when the latter equals the static stall angle.

To evaluate whether the effective angle of attack accounts for the observed reaction delay in reattachment onset, we recalculated the reaction delay timing using the effective angle of attack.
This means that we measured the time difference between the moment the effective angle of attack reached the critical stall angle of \ang{21.4}, and the moment the leading-edge suction parameter reaches the critical value $\kindex{A}{0}^*$. 
When considering the effective angle of attack, the reaction delay is slightly higher than the reaction delay defined based on the geometric angle of attack, as the effective angle is lower than the geometric angle during the pitch down motion. 
The shift between the reaction state timings based on the geometric or the effective angle of attack is of the order of one convective time. 
Even though the effective angle of attack offset increases with increasing effective pitch rate, the time it takes to cover the angle of attack offset remains approximately constant, and the effect on the reaction time delay is an approximately constant shift over the range of motions analysed here. 
The overall decay of the reaction time delay as a function of the effective pitch rates is not the result of an effective angle of attack variation induced by the pitching motion. 

\section{Statistical analysis to determine the critical lading-edge suction parameter}\label{appendix:A0statistical}
The critical leading-edge suction parameter at recovery was determined through statistical analysis of the experimental data.
The leading-edge suction parameter was calculated from the experimentally measured surface pressure distribution around the leading edge. 
The leading edge suction is inherently low when the flow is separated and experiences significant cycle-to-cycle variations which leads to noise and significant fluctuations in the data which complicate the extraction of threshold value that triggers flow recovery. 

We propose here a robust procedure to determine a value of $\kindex{A}{0}$ for each pitch rate beyond which we consistently observe stall recovery in our experiments that can handle outliers in the data.
The analysis uses the data from multiple oscillation cycles that are recorded for each pitch rate.
The number of measured pitching cycles ranges from \num{39} for the lowest pitching frequency to \num{80} for the highest pitching frequency. 
Each cycle was treated as an individual observation in the population corresponding to a given case. 
We systematically loop through different candidate threshold values of $\kindex{A}{0}$ and assess its threshold sufficiency. 
A threshold value is considered sufficient if the suction parameter exceeds it only once at the end of the recovery range, without fluctuating around it.
Any drop in the suction parameter after exceeding a candidate threshold value would indicate a return to the separated flow state, suggesting the threshold was insufficient to trigger flow recovery.
For each pitch rate case, we count the relative number of cycles for the which the threshold sufficiency criterion is met for different values of $\kindex{A}{0}$. 
We refer to this relative number of cycles as our success rate $f$.

\begin{figure}
\centering
\includegraphics{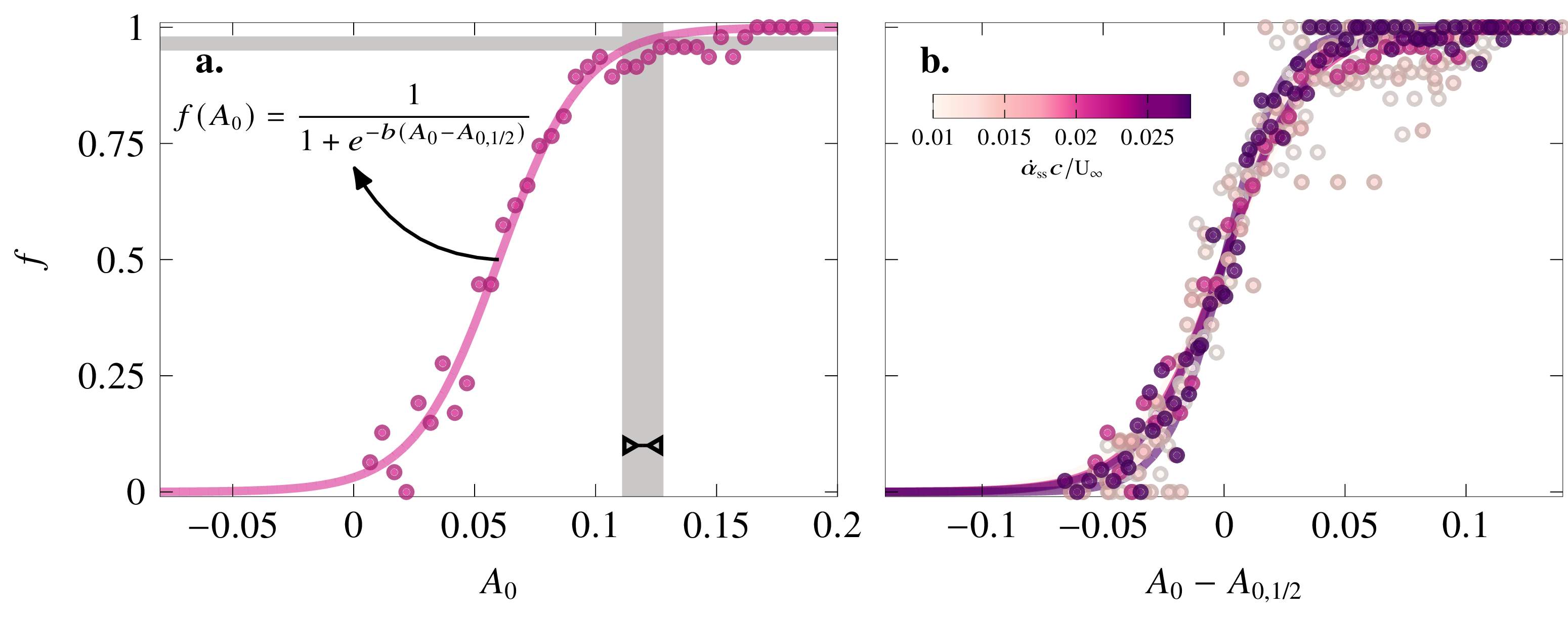}
\caption{(a) Success rate as a function of the candidate threshold value of the leading-edge suction parameter $\kindex{A}{0}$, fitted with the sigmoid model given by \cref{eq:sigmoid}. 
The horizontal shaded band shows the target success range between \numrange[range-phrase = {\text{\ and\ }}]{0.95}{0.98}, and the vertical band highlights the corresponding range of critical $\kindex{A}{0}$ values.
(b) Summary of the threshold sufficiency success rate distribution and sigmoidal curve fit for all tested pitch rates.}
\label{fig:sigmoids}
\end{figure}

The typical distribution of the success rate as a function of the candidate threshold value of $\kindex{A}{0}$ is presented in \cref{fig:sigmoids}a for a pitching motion with $\kindex{\dot{\alpha}}{ss}=\num{0.0208}$.  
The success rate increases with increasing threshold values following a distribution described by a dose-response model, which describes how the likelihood of a response increases with increasing exposure or dose.
We model this trend by fitting a sigmoidal relationship to the success rate in the form of:
\begin{equation}\label{eq:sigmoid}
	f(\kindex{A}{0}) = \frac{1}{1 + e^{-b(\kindex{A}{0} - \kindex{A}{0,1/2})}},
\end{equation}
where $\kindex{A}{0}$ is the candidate threshold leading-edge suction value, $b$ controls the steepness of the curve and is a fitting parameter, and $\kindex{A}{0,1/2}$ is the inflection point which represents the threshold at which the success rate reaches \qty{50}{\percent}.
The sigmoid shape reflects that success becomes more probable as the threshold increases and describes well the response for all pitch rates (\cref{fig:sigmoids}b).
Based on the fitted response curves, we then extracted the range of $\kindex{A}{0}$ thresholds corresponding to success rates between \numrange[range-phrase = {\text{\ and\ }}]{0.95}{0.98} (\cref{fig:sigmoids}a).
This procedure was repeated for each pitch rate (\cref{fig:sigmoids}b), and the resulting critical suction parameter ranges were obtained for all cases (\cref{fig:A0_critical}).

\section{Effect of integration range on the critical leading-edge suction parameter}\label{appendix:LESPintegral}
\begin{figure}
\centering
\includegraphics{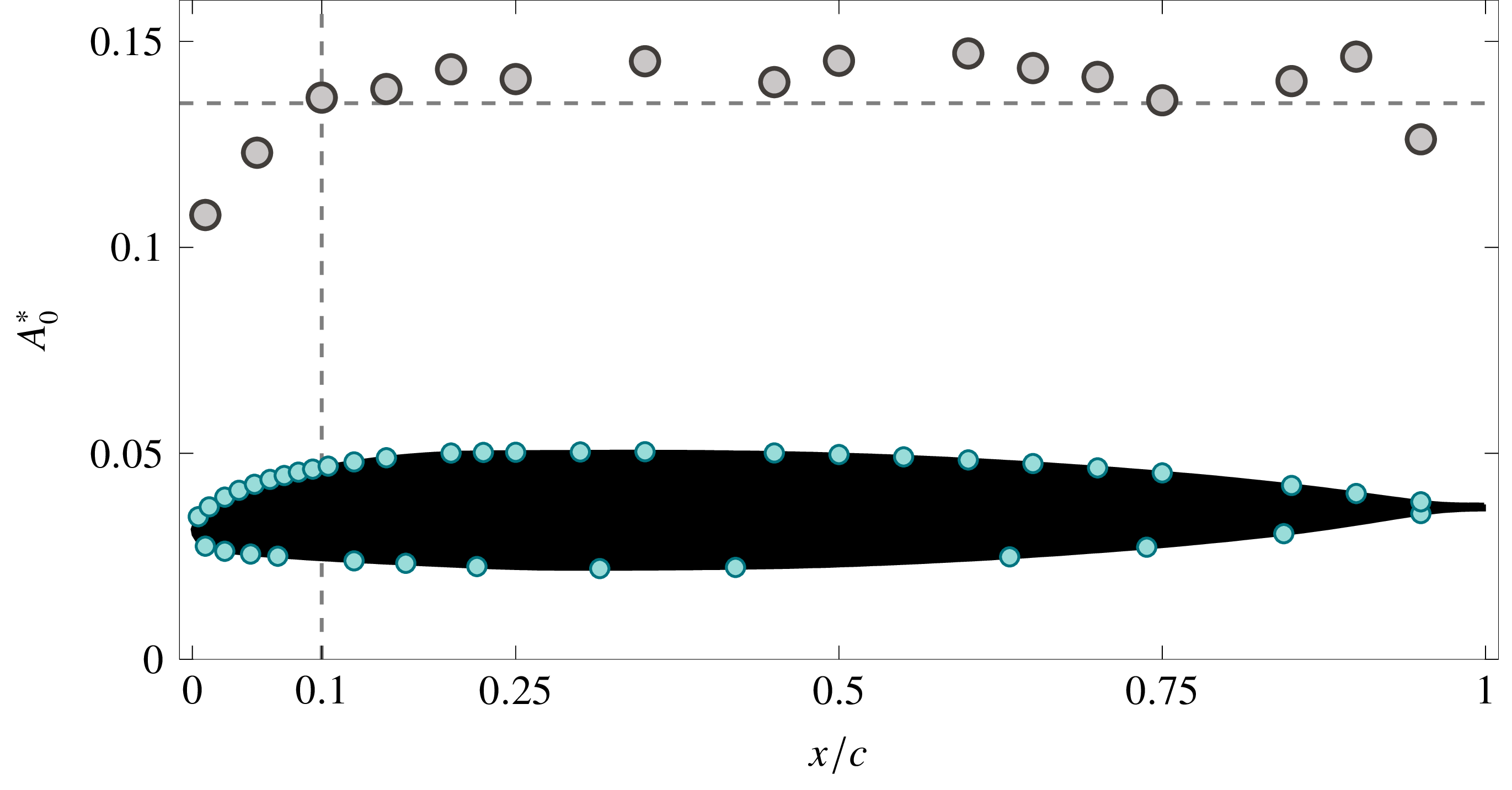}	
\caption{Critical leading-edge suction parameter value as a function of the maximum chordwise locations taking into account for the integration of the leading-edge suction parameter. 
The location of the pressure sensor are indicated on the airfoil profile. 
The horizontal dashed line indicated the critical leading-edge suction parameter value ($\kindex{A}{0}^*$) obtained using the first \qty{10}{\percent} of the chord.}
\label{fig:LEarea}	
\end{figure}

The magnitude of the leading-edge suction parameter depends on the chord-wise distance over which the pressure data are integrated. 
In the main part of the paper, we use the pressure data from the sensors in the first \qty{10}{\percent} of the chord to determine the leading-edge suction parameter. 
The magnitude of the leading-edge suction parameter will affect the value of the critical leading-edge suction parameter ($\kindex{A}{0}^*$). 
Here, we examine how the integration range affects the value of the critical leading-edge suction parameter.
We varied the integration ranges from the front \qty{1}{\percent} to the full chord length and follow the procedures outlined in \cref{sec:criticalA0} and \cref{appendix:A0statistical} to obtain the critical leading edge suction parameter for all pitching motions as a function of the integration range (\cref{fig:LEarea}). 

The most significant change in the critical values occurs when the integration range is increased from \qty{1}{\percent} to \qty{10}{\percent}. 
When the integration range is extended beyond the first \qty{10}{\percent}, the critical leading-edge suction parameter reaches a plateau and fluctuates between \num{0.13} and \num{0.15}. 
The critical value presented in \cref{fig:LEarea} is the maximum of the critical values obtained for the individual pitching motions. 
The variations are mostly attributed to variations in the quality of the sigmoid model fits for the different kinematics (\cref{appendix:A0statistical}).
   
\bibliographystyle{jfm}
\bibliography{mainbib}
	
\end{document}